\documentclass[11pt]{article}

\usepackage{amsmath}
\usepackage{graphicx}
\usepackage{amsfonts}
\usepackage{amssymb}
\usepackage{epsfig}
\usepackage{color}
\usepackage{psfrag}
\usepackage{epstopdf}

\newcommand{\EQ}{\begin{equation}}
\newcommand{\EN}{\end{equation}}
\newcommand{\be}{\begin{equation}}
\newcommand{\ee}{\end{equation}}
\newcommand{\bea}{\begin{eqnarray}}
\newcommand{\eea}{\end{eqnarray}}

\begin{document} \setcounter{page}{0}
\topmargin 0pt
\oddsidemargin 5mm
\renewcommand{\thefootnote}{\arabic{footnote}}
\newpage
\setcounter{page}{0}
\topmargin 0pt
\oddsidemargin 5mm
\renewcommand{\thefootnote}{\arabic{footnote}}
\newpage
\begin{titlepage}
\vspace{0.5cm}
\begin{center}
{\large {\bf Particles, string and interface}\\
{\bf in the three-dimensional Ising model}}\\
\vspace{1.8cm}
{\large Gesualdo Delfino$^{1,2}$, Walter Selke$^3$ and Alessio Squarcini$^{4,5}$}\\
\vspace{0.5cm}
{\em $^1$SISSA -- Via Bonomea 265, 34136 Trieste, Italy}\\
{\em $^2$INFN sezione di Trieste, 34100 Trieste, Italy}\\
{\em $^3$Institute for Theoretical Physics, RWTH Aachen University, 52056 Aachen, Germany}\\
{\em $^4$Max-Planck-Institut f\"ur Intelligente Systeme, 
Heisenbergstr. 3, D-70569, Stuttgart, Germany}\\
{\em $^5$IV. Institut f\"ur Theoretische 
Physik, Universit\"at Stuttgart, Pfaffenwaldring 57, D-70569 
Stuttgart, Germany}\\
\end{center}
\vspace{1.2cm}

\renewcommand{\thefootnote}{\arabic{footnote}}
\setcounter{footnote}{0}

\begin{abstract}
\noindent
We consider the three-dimensional Ising model slightly below its critical temperature, with boundary conditions leading to the presence of an interface. We show how the interfacial properties can be deduced starting from the particle modes of the underlying field theory. The product of the surface tension and the correlation length yields the particle density along the string whose propagation spans the interface. We also determine the order parameter and energy density profiles across the interface, and show that they are in complete agreement with Monte Carlo simulations that we perform. 
\end{abstract}
\end{titlepage}

\newpage
\section{Introduction}
The notion of interface plays an important role in different areas of physics. In statistical systems, the separation of different phases is characterized through the formation of an interface. In particle physics, the simplest description of confinement is in terms of a flux tube (a string) that connects the quarks and whose time propagation spans an interface. Lattice discretization establishes a direct connection between the two problems when duality relates a spin model to a lattice gauge theory, with the Ising model providing the basic example \cite{ID}. Effective descriptions adopting interfacial fluctuations as the basic degrees of freedom result into capillary wave theory \cite{BLS} on one side, and effective string actions \cite{Goto,Nambu} on the other.

In this paper we consider the three-dimensional Ising model in its scaling limit below the critical temperature $T_c$, where it is described by field theory, and use the asymptotic particle states of the bulk field theory as the basis on which to perform expansions in momentum space. Introducing boundary states that induce the presence of an interface, the formalism allows us to determine the interfacial properties, including the magnetization and energy density profiles at leading order in the linear size $R$ of the interface. We then numerically determine the profiles through Monte Carlo simulations for different values of the temperature $T$ and of the size $R$, and exhibit complete agreement with the analytic results, in absence of adjustable parameters.

The paper is organized as follows. In the next section, we introduce the boundary state setup and use it to determine the interfacial free energy and the expression of one-point functions, from which we then obtain the magnetization and energy density profiles. Section~3 is devoted to Monte Carlo simulations of the near-critical Ising model on the cubic lattice and to comparison with the analytic results for the profiles. Finally, in section~4 we discuss several implications of our results and point out lines of further development.

\section{From particles to the interface}
We consider the Ising model with reduced Hamiltonian 
\EQ
{\cal H}=-\frac{1}{T}\sum_{<i,j>}{s}_i{s}_j\,,
\label{H}
\EN
where ${s}_i=\pm 1$ is the spin variable located at the site $i$ of a cubic lattice, and the sum is performed over all pairs of nearest neighboring sites. We focus on the case of temperatures $T<T_c$, in which the spin reversal $Z_2$ symmetry of the Hamiltonian is spontaneously broken, i.e. 
\EQ
M\equiv|\langle{s}_i\rangle|\neq 0\,;
\EN
as usual, $\langle\cdots\rangle$ denotes the average over spin configurations weighted by $e^{-{\cal H}}$. More precisely, we restrict our attention to the temperature range slightly below $T_c$, where the correlation length $\xi$ becomes large and the system is described by a three-dimensional Euclidean field theory, which in turn is the continuation to imaginary time of a quantum field theory in $(2+1)$ dimensions. This amounts to consider the scaling region below $T_c$, and our analytic results quantitatively hold as long as the temperature dependence of the observables is ruled by the Ising critical exponents. As we detail in section~3, this scaling regime is distant above the roughening transition temperature $T_r$ \cite{WGL} below which the fluctuations of the interface are suppressed.  In the continuum we will denote by $r=(x,y,z)$ a point in Euclidean space, $z$ being the imaginary time direction, and by ${s}(r)$ the spin field. We refer to this translationally and rotationally invariant theory as the {\it bulk} theory.

\begin{figure}[t]
\begin{center}
\includegraphics[width=9cm]{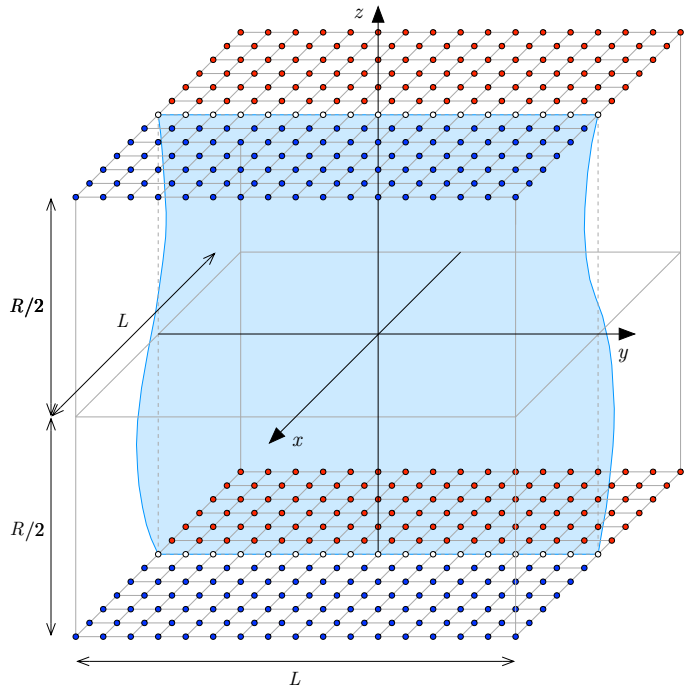}
\caption{Geometry considered for the Ising model below $T_c$, with $L\to\infty$ in the theoretical analysis. Boundary spins on the top and bottom surfaces are fixed to $1$ (red) for $x<0$ and to $-1$ (blue) for $x>0$, and left free for $x=0$, so that an interface (one configuration is shown) runs between the axes $x=0$ on these surfaces.}
\label{geometry}
\end{center}
\end{figure}

We then focus on the case in which the system is finite in the $z$ direction, with $z\in(-R/2,R/2)$ and $R\gg\xi$, while the size in the $x$ and $y$ directions is kept infinite in the theoretical analysis. The boundary conditions at $z=\pm R/2$ are chosen in such a way that $s_i=1$ for $x<0$ and $s_i=-1$ for $x>0$; the spins are left unconstrained for $x=0$. It follows that for $z=0$ and $R$ large, the magnetization $\langle s(r)\rangle_{+-}$ tends to the bulk value $M$ as $x\to-\infty$, and to $-M$ as $x\to\infty$; we denote by $\langle\cdots\rangle_{+-}$ configurational averages with the boundary conditions we have fixed. The two pure phases for $x$ large and negative and $x$ large and positive are separated around $x=0$ by an interfacial region spanned by the fluctuations of an interface running between the straight lines $x=0$ at $z=\pm R/2$ (Figure~\ref{geometry}). It is our goal to determine the expectation value $\langle\Phi(x,y,0)\rangle_{+-}$ of a field $\Phi(r)$. 

The fact that the scaling region around the critical temperature is described by a field theory is well known and widely used, in particular for the perturbative determination of the Ising critical exponents \cite{ID}. On the other hand, a field theory admits a particle description (see e.g. \cite{Ryder}), and it is this description that we will exploit for our study of the interface. Non-translationally invariant states of the system correspond to field theoretical states with nonzero energy and momentum. Energy and momentum are carried by the particles of the bulk field theory\footnote{It is worth stressing that the particles we refer to throughout the paper describe the collective excitation modes of the system, and should not be confused with the individual molecules of a fluid whose near-critical properties are described by the field theory.}. They evolve in two spatial dimensions (the $x$ and $y$ directions of Figure~\ref{geometry}) and one imaginary time dimension (the $z$ direction). The analytic continuation to imaginary (or Euclidean) time $z=it$ is the usual way \cite{ID,Ryder} to exploit the fact that a near-critical statistical system at thermal equilibrium in $d$ spatial dimensions can be mapped onto a quantum system in $d-1$ spatial dimensions and one time dimension. In our case $d=3$, and the rotational invariance (isotropy) of the statistical system in three Euclidean dimensions is mapped into relativistic invariance of the quantum system in $(2+1)$ dimensions. It follows that the energy $E_{\bf p}$ of a particle mode with momentum ${\bf p}=(p_x,p_y)$  and mass $m$ obeys the relativistic dispersion relation $E_{\bf p}=\sqrt{{\bf p}^2+m^2}$. The asymptotic $n$-particle states $|{\bf p}_1,{\bf p}_2,\ldots,{\bf p}_n\rangle$ of the bulk field theory provide a basis on which generic excitations of the system can be expanded. They are eigenstates of the energy and momentum operators with eigenvalues $\sum_{i=1}^nE_{{\bf p}_i}$ and $\sum_{i=1}^n{{\bf p}_i}$, respectively. 

The boundary conditions that we impose at $z=\pm R/2$ correspond in the field theory to boundary states $|B(\pm R/2)\rangle=e^{\pm\frac{R}{2}H}|B(0)\rangle$ of the Euclidean time evolution, with $H$ denoting the energy operator (Hamiltonian) of the $(2+1)$-dimensional quantum system. A boundary state can be expanded on the basis of asymptotic states of the bulk field theory. For our boundary conditions below $T_c$, the boundary states correspond to an excitation (a string) extending for all values of $y$, and whose propagation in the $z$ direction spans the interface. It follows that the number of particles entering the states in the expansion has to be extensive in the $y$ direction, and is therefore infinite. In order to regulate our expressions, we write this number as $N\propto L\to\infty$, and this limit will be understood in the following. We then write 
\EQ
|B(\pm R/2)\rangle=\frac{1}{\sqrt{N!}}\int\prod_{i=1}^N\frac{d{{\bf p}_i}}{(2\pi)^2E_{{\bf p}_i}}\,f({\bf p}_1,\ldots,{\bf p}_N)\,e^{\pm\frac{R}{2}\sum_{i=1}^NE_{{\bf p}_i}}\,\delta\left(\sum_{i=1}^N p_{y,i}\right)\,|{\bf p}_1,\ldots,{\bf p}_N\rangle+\ldots\,,
\label{B}
\EN
where $f({\bf p}_1,\ldots,{\bf p}_N)$ is an amplitude, particle states are normalized as $\langle{\bf p}'|{\bf p}\rangle=(2\pi)^2E_{\bf p}\,\delta({\bf p}-{\bf p}')$, and the delta function enforces translation invariance in the $y$ direction. $m$ is the mass of the lightest particle in the spectrum of the spontaneously broken phase of the bulk field theory. It enters the large distance decay of the spin-spin correlator as $\langle s(r)s(0)\rangle\sim e^{-m|r|}$. Comparison with the definition of the correlation length yields
\EQ
\xi=1/m\,.
\label{xi}
\EN
States involving heavier particles also enter the expansion (\ref{B}) in the part that we do not write explicitly. As we will immediately discuss, they produce only subleading corrections in the large $R$ limit we are interested in. 

The partition function $Z_{+-}$ corresponding to our boundary conditions is given by the overlap between the two boundary states, which implements the sum over configurations of particles propagating between the bottom and top surfaces. Then we have 
\bea
Z_{+-} & = & \langle B(R/2)|B(-R/2)\rangle=\langle B(0)|e^{-RH}|B(0)\rangle\nonumber\\
&\sim & \frac{L}{2\pi}|f_0|^2\int\prod_{i=1}^N\frac{d{\bf p}_i}{(2\pi)^2 m}\,\delta\left(\sum_{i=1}^N p_{y,i}\right)\,e^{-R\left(Nm+\sum_{i=1}^N\frac{{\bf p}_i^2}{2m}\right)}\nonumber \\
&=& \frac{L|f_0|^2e^{-RNm}}{(2\pi)^{2(N+1)}}\left(\frac{2\pi}{R}\right)^N\sqrt{\frac{2\pi R}{Nm}}\,\,,
\label{Z}
\eea
where we used the fact that the large $R$ limit forces all momenta to be small, defined $f_0=f(0,\ldots,0)$, exploited $2\pi\delta(p)=\int e^{iup}du$, and regularized $\delta(0)$ as $L/2\pi$, so that here and in the following formulae $L\to\infty$ is the size of the system in the $y$ direction. Here and below the symbol $\sim $ indicates omission of terms subleading for large $R$. It appears from (\ref{Z}) how the contribution to $Z_{+-}$ of a state in which a particle of mass $m$ is replaced by one of mass $m'>m$ is further suppressed at large $R$ by a factor $e^{-(m'-m)R}$.

The interfacial free energy, i.e. the contribution to the free energy due to the presence of the interface, is $F_{\textrm{interface}}=-\ln Z_{+-}$. The interfacial tension $\sigma$ is defined as the interfacial free energy per unit area, $F_{\textrm{interface}}/LR$, for both $L$ and $R$ going to infinity. Hence, it follows from (\ref{Z}) and (\ref{xi}) that it is given by\footnote{Since the limit $L\to\infty$ is understood, in (\ref{tension}) we only indicate the limit $R\to\infty$.}
\EQ
\sigma=-\lim_{R\to\infty}\frac{1}{LR}\ln Z_{+-}=\kappa\,m^2=\frac{\kappa}{\xi^2}\,,
\label{tension}
\EN
where
\EQ
\kappa=\frac{N\xi}{L}\,.
\label{kappa}
\EN
The reason for introducing $\kappa$ is that, being dimensionless, it is a universal number, namely a number that near criticality is the same for different lattice discretizations. It also follows that $N/L$, the number of particles per unit length along the string, can be written as $N/L=\sigma\xi$; equivalently, there are $\kappa$ particles per correlation length in the $y$ direction. Notice that, since the energy of the state is the sum of the particle energies, in (\ref{B}) the interaction among the particles is taken into account by the amplitude $f({\bf p}_1,\ldots,{\bf p}_N)$. In the large $R$ limit that we consider this function is projected to the constant $f_0$, which only corresponds to the arbitrary normalization of the boundary state and can be set, in particular, to one. We deduce that the large $R$ limit is one of weakly interacting, and then (in average) widely separated, particles. This conclusion fully agrees with the fact that the known Monte Carlo value $\kappa=0.1084(11)$ \cite{CHP} corresponds to an average interparticle distance in the $y$ direction of about ten correlation lengths. It is particularly interesting that the particle description provides insight on a measurable and universal quantity like $\kappa$. 

Notice also that, while $N$ and $L$ enter our formulae as regulators that go to infinity, measurable quantities like (\ref{tension}) only depend on the finite ratio (\ref{kappa}). This internal consistency of the theory is further illustrated by the one-point functions (i.e. expectation values of local observables) that we now compute. It is also worth stressing how, since the initial expression (\ref{B}) includes all fluctuations (sum over all particle excitations and all momenta), the large $R$ asymptotics that we derive are exact.

The one-point functions at $z=0$ are given by
\bea
\hspace{-1cm}G_\Phi(x)&\equiv&\langle\Phi(x,y,0)\rangle_{+-}=\frac{1}{Z_{+-}}\,\langle B(R/2)|\Phi(x,y,0)|B(-R/2)\rangle \nonumber\\
&\sim &\frac{|f_0|^2}{Z_{+-}N!}\int\prod_{i=1}^N\left(\frac{d{\bf p}_i}{(2\pi)^2 m}\frac{d{\bf q}_i}{(2\pi)^2 m}\right)\,\delta\left(\sum_{i=1}^N p_{y,i}\right)\delta\left(\sum_{i=1}^N q_{y,i}\right)\nonumber\\
&\times & F_\Phi({\bf p}_1,\ldots,{\bf p}_N|{\bf q}_1,\ldots,{\bf q}_N)\,e^{-\frac{R}{2}\left(2Nm+\sum_{i=1}^N\left(\frac{{\bf p}_i^2}{2m}+\frac{{\bf q}_i^2}{2m}\right)\right)+ix\sum_{i=1}^N\left(p_{x,i}-q_{x,i}\right)},
\label{vPhi0}
\eea
where we again consider the large $R$ limit, the vanishing of the $y$ component of the total momentum yields $y$-independence, and the matrix element
\bea
&& \hspace{-.5cm}F_\Phi({\bf p}_1,\ldots,{\bf p}_N|{\bf q}_1,\ldots,{\bf q}_N)=\langle{\bf p}_1,\ldots,{\bf p}_N|\Phi(0)|{\bf q}_1,\ldots,{\bf q}_N\rangle\label{ff}\\
&& \hspace{-.5cm}=\langle{\bf p}_1,\ldots,{\bf p}_N|\Phi(0)|{\bf q}_1,\ldots,{\bf q}_N\rangle_c+(2\pi)^2m\,\delta({\bf p}_1-{\bf q}_1)\langle{\bf p}_2,\ldots,{\bf p}_N|\Phi(0)|{\bf q}_2,\ldots,{\bf q}_N\rangle_c+\ldots\nonumber
\eea
is evaluated for small momenta. In the second line we take into account its decomposition in connected and disconnected parts, the latter originating from annihilation of particles on the left with particles on the right \cite{Ryder}; the subscript $c$ denotes connected matrix elements, and the dots indicate that all possible annihilations have to be included. It follows from (\ref{vPhi0}) that each power of momentum in the integral contributes a factor $R^{-1/2}$ to the one-point function. Since each annihilation in (\ref{ff}) produces a delta function $\delta({\bf p}_i-{\bf q}_j)$, and then a factor $R$, the leading contribution to (\ref{vPhi0}) for large $R$ is obtained maximizing the number of annihilations. Since $N$ annihilations leave an $x$-independent term $C_\Phi$, the interesting term is that with $N-1$ annihilations. Taking also into account that there are $N!N$ ways of performing $N-1$ annihilations, we finally obtain
\EQ
G_\Phi(x)\sim C_\Phi+\frac{\kappa R}{(2\pi)^2m}\int d{\bf p}d{\bf q}\,\delta(p_y-q_y) \,F_\Phi^c({\bf p}|{\bf q})\,e^{-\frac{R}{4m}({\bf p}^2+{\bf q}^2)+ix(p_x-q_x)}\,.
\label{Phi}
\EN
If $F_\Phi^c({\bf p}|{\bf q})\equiv\langle{\bf p}|\Phi(0)|{\bf q}\rangle_c$ behaves as momentum to the power $\alpha_\Phi$, the $x$-dependent part of (\ref{Phi}) behaves as
\EQ
R^{-(1+\alpha_\Phi)/2}\,.
\label{Rdep}
\EN
We also have that the integral term in (\ref{Phi}) is even (resp. odd) in $x$ when $F_\Phi^c({\bf p}|{\bf q})|_{p_y=q_y}$ is even (resp. odd) under exchange of $p_x$ and $q_x$. 

The fact that the magnetization profile $G_s(x)$ has to be an odd function of $x$ interpolating between $M$ and $-M$ fixes $C_s=0$ and $\alpha_s=-1$. This leads to\footnote{A suitable extension of (\ref{Fs}) to generic small momenta appears to be $F_s^c({\bf p}|{\bf q})=c_s \bigl[({\bf p}-{\bf q})^{2}\bigr]^{-1/2}$. For $q_y=p_y$ it yields $c_s/\sqrt{(p_x-q_x)^{2}}$, and (\ref{Fs}) is the way of extracting the sign from the square root compatible with the usual analyticity requirements \cite{ELOP} for the matrix elements, which do not allow for absolute values.}
\EQ
F_s^c({\bf p}|{\bf q})|_{p_y=q_y}=\frac{c_s}{p_x-q_x}\,, \hspace{1cm}p_x,q_x\to 0\,.
\label{Fs}
\EN
Upon insertion in (\ref{Phi}) the pole in $p_x-q_x$ is conveniently canceled by differentiation with respect to $x$. Performing the momentum integrations and integrating back in $x$ we obtain
\EQ
G_s(x)\sim-M\,\textrm{erf}(\eta)\,,
\label{mag_profile}
\EN
\EQ
\eta = \sqrt{\frac{2}{R\xi}}\,x \,,
\label{eta}
\EN
and $c_s=-2iM/\kappa$. The error function entering the magnetization profile (\ref{mag_profile}) already appears in the exact result in two dimensions \cite{Abraham,Abraham_review,DV}, a circumstance that we will discuss in section~4.

The energy density profile $G_\varepsilon(x)$ has to be an even function of $x$, but the value of $\alpha_\varepsilon$ is not obvious a priori and remains as a parameter. We then write
\EQ
F_\varepsilon^c({\bf p}|{\bf q})=c_{\varepsilon} \bigl[({\bf p}+{\bf q})^{2}\bigr]^{\alpha_\varepsilon/2}\,, \hspace{1cm}{\bf p},{\bf q}\to 0\,. 
\label{Feps}
\EN
The integrations in (\ref{Phi}) are easily performed passing to the variables ${\bf p}\pm{\bf q}$ and yield the result
\EQ
G_\varepsilon^c(x)\equiv G_\varepsilon(x)-C_\varepsilon\sim \frac{b_\varepsilon\,\xi^{-X_\varepsilon}}{(R/\xi)^{(1+\alpha_\varepsilon)/2}}\,e^{-\eta^2}\,,
\label{eps_profile}
\EN
where we exploited the fact that the result must have the scaling dimension $X_\varepsilon$ of the energy density field to express the temperature dependence of the prefactor of the Gaussian in terms of the correlation length; $b_\varepsilon$ is then a dimensionless constant depending on the normalization of $\varepsilon(x)$. Equation (\ref{eta}) shows that the width of the Gaussian in (\ref{eps_profile}), i.e. the width of the interfacial fluctuations around the pinning position $x=0$, is infinite for $R=\infty$. This accounts for the vanishing of the magnetization profile (\ref{mag_profile}) for $R=\infty$: due to the infinite fluctuation width, the interface can be found with equal probability to the right or to the left of any point along the $x$-axis, and the average yields a zero magnetization. However, for $R$ finite, no matter how large, translation invariance along the $x$-axis is broken.

\section{Comparison with Monte Carlo simulations}
We now compare the theoretical predictions with Monte Carlo simulations of the Ising model on the simple cubic lattice. Most of the numerical data for the bulk quantities entering our analysis are given, for example, in \cite{CH} with an accuracy sufficient for our purposes. In particular, we have $1/T_c=0.2216544(3)$ (corresponding to $T_c\simeq 4.51153$), $\nu=1/(3-X_\varepsilon)=0.6310(15)$, $\beta=0.3270(6)$. The critical exponents $\nu$ and $\beta$ rule the behavior of the correlation length and spontaneous magnetization for $T\to T_c^-$ as (see e.g. \cite{ID})
\bea
\xi &\simeq & \xi_0\,(T_c-T)^{-\nu}\,,
\label{xi_scaling}\\
M &\simeq & B\,(T_c-T)^\beta\,,
\label{mag_scaling}
\eea
respectively. The critical amplitude $\xi_0$ can be obtained from a fit of the data listed in Table~3 of \cite{CH} and reads $\xi_0\simeq 0.668$. For the bulk magnetization, the numerical approximation \cite{TB}
\EQ
M \simeq t^{0.32694} (1.6919-0.34357\,t^{0.50842}-0.42572\,t)\,
\label{M_corr}
\EN
is available, which also estimates the first corrections to (\ref{mag_scaling}) for small $t=(T_c-T)/T_c$ and fits very well the data in the temperature range of our interest \cite{CH,TB}. 

We shall focus on the numerical determination, by Monte Carlo techniques, of the profiles for the magnetization and the energy density for which we derived the analytic expressions (\ref{mag_profile}) and (\ref{eps_profile}). The system is simulated on the simple cubic lattice in the volume $x\in(-L/2,L/2)$, $y\in(-L/2,L/2)$, $z\in(-R/2,R/2)$, with $L$ sufficiently larger than $R$ in order to take into account that we want to compare with theoretical results corresponding to infinite $L$. The boundary spins are fixed as previously described for $z=\pm R/2$, and are left free on the other boundaries. 

\begin{figure}[t]
\begin{center}
\includegraphics[width=10cm]{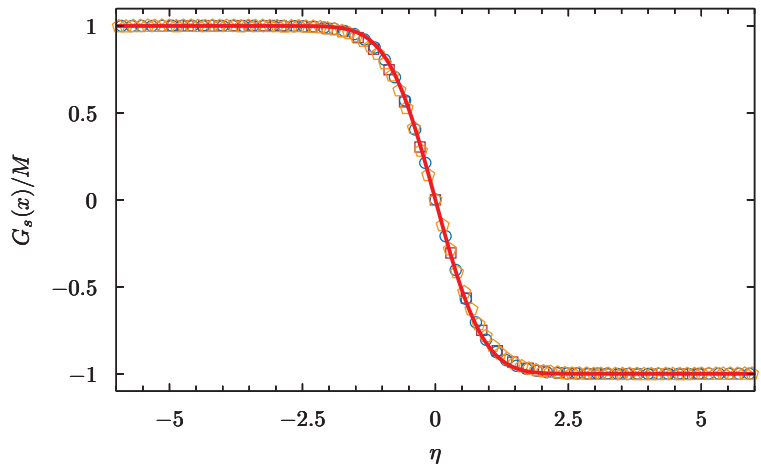}
\caption{Analytic result (\ref{mag_profile}) for the magnetization profile (continuous curve) and the corresponding Monte Carlo results (data points). The latter are obtained for $T=4.2$, $R=17$, $L=55$ (squares), $T=4.3$, $R=31$, $L=91$ (circles), and $T=4.4$, $R=41$, $L=121$ (pentagons). The scaling variable $\eta$ is given by (\ref{eta}).
}
\label{mag_plot}
\end{center}
\end{figure}

\begin{figure}[t]
\begin{center}
\includegraphics[width=10cm]{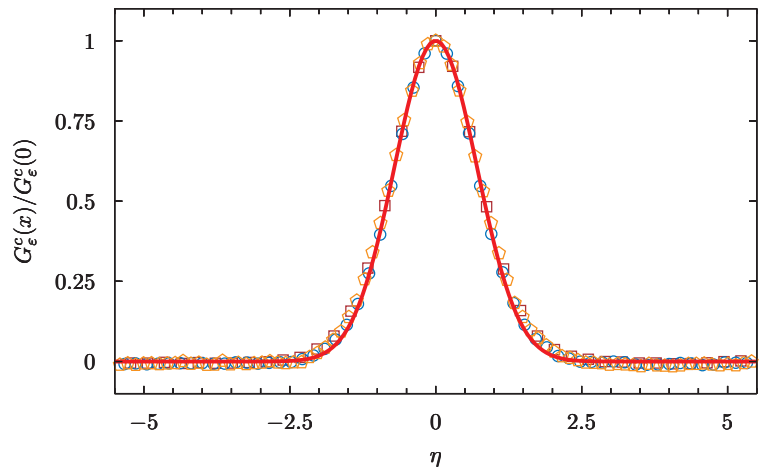}
\caption{Analytic result (\ref{eps_profile}) for the energy density profile (continuous curve) and the corresponding Monte Carlo results (data points). The data symbols refer to the same temperatures and sizes as in Figure~\ref{mag_plot}.
}
\label{eps_plot}
\end{center}
\end{figure}

As in our recent Monte Carlo simulations for two-dimensional Potts models \cite{DSS1} and three-dimensional XY model \cite{DSS2}, the standard Metropolis algorithm \cite{LB} turned out to be useful. In particular, to test the predictions of the theory and to study finite size effects, we varied the lattice sizes and the temperature. The linear dimension $R$ ranged from 11 to 47, with $L$ ranging from 55 to 121 (the lengths are expressed in units of the lattice spacing). Data were taken at temperatures above the roughening transition, $T_r \simeq 2.45$ (see \cite{HP}), and below $T_c\simeq 4.51$ of the Ising model on the cubic lattice, concentrating on the region $4.1\lesssim T<T_c$, where the bulk correlation length shows the scaling behavior (\ref{xi_scaling}). This is the scaling region in which the Monte Carlo results can be compared with our analytical results. Specifically, we analyzed the temperature interval between $T= 4.2$ and 4.4. As usual, to obtain numerical results of high quality, we varied the length of the Monte Carlo runs, in between $10^5$ and $5\cdot 10^7$ Monte Carlo steps per site (MCS). Studying lattices of finite size below the critical point, we then performed simulations with $10^7$ MCS. Thermal averages were taken for the quantities of interest of the theory, the magnetization and energy density profiles in the center of the lattice. To test and determine the accuracy of the simulation data, we averaged over, at least, four independent Monte Carlo runs, using different random numbers in each realization. The resulting error bars normally did not exceed the size of the symbols in Figures~\ref{mag_plot} and \ref{eps_plot}, where final Monte Carlo results together with the theoretical predictions are shown.

The magnetization and the energy density are local observables and their determination below $T_c$ can be ordinarily performed as in the bulk case (see \cite{CH}). The difference in our case is that the boundary conditions that we adopt induce the $x$-dependence that we determined in (\ref{mag_profile}) and (\ref{eps_profile}) starting from the particle description of the interface. The simulations are necessarily performed for $L$ finite, but for $L$ sufficiently larger than $R$ the Monte Carlo data are expected to reproduce the infinite $L$ analytical results (\ref{mag_profile}) and (\ref{eps_profile}), in which the profiles flatten on the constant bulk values for $|x|$ large. This is fully confirmed by the comparisons between theory and data in Figures~\ref{mag_plot} and~\ref{eps_plot}.

The profiles are determined along the axis $y=z=0$, with $|x|$ sufficiently far from the boundaries. Figure~\ref{mag_plot} shows that the Monte Carlo data that we obtain for the magnetization for different values of $T$ and $R$ exhibit the theoretically predicted collapse on a single curve once divided by $M$ and plotted as a function of the scaling variable (\ref{eta}). While the observation of this scaling behavior is in itself a notrivial confirmation of the theory, the figure also shows that the numerically determined profile agrees very well with the analytical result $-\textrm{erf}(\eta)$, see (\ref{mag_profile}). It is worth stressing that the comparison contains no adjustable parameter. 

For the energy density, which on the lattice corresponds to $\varepsilon_i=\sum_{j\sim i}{s}_i{s}_j$, with the sum running over the nearest neighbors of site $i$, we consider the profile $G^c_\varepsilon(x)$, which we obtain subtracting the plateau (bulk) value that we read from the data. Figure~\ref{eps_plot} shows that the Monte Carlo data for $G^c_\varepsilon(x)/G^c_\varepsilon(0)$ exhibit the expected collapse when plotted against $\eta$; agreement with the analytic result $e^{-\eta^2}$ is also very good, again without free parameters. 

It is worth stressing that, as confirmed by the comparison with Monte Carlo data in Figures~\ref{mag_plot} and \ref{eps_plot}, the results (\ref{mag_profile}) and (\ref{eps_profile}) are the answer to the specific problem that we studied, namely that of temperatures in the scaling region below $T_c$ and interpinning distance $R$ as the only finite size variable. These specifications correspond to the goal of this paper: describing the near-critical system with an interface starting from the particle modes of the bulk field theory, and doing so in an analytically exact way that allows for a parameter-free comparison with Monte Carlo simulations of the system on a lattice. 
Different system specifications are expected to lead to expressions for the profiles qualitatively similar to (\ref{mag_profile}) and (\ref{eps_profile}) from the point of view of the $x$-dependence, but differing from them in the functional form and/or parameter dependence.

\section{Discussion}
In this paper we have considered the three-dimensional Ising model slightly below the critical temperature $T_c$, with boundary conditions enforcing the presence of an interface running between two straight lines separated by a distance $R$ much larger than the bulk correlation length $\xi$. We have shown analytically how the interface emerges from the study of the bulk field theory supplemented with the required boundary conditions. In particular, we showed how the string whose imaginary time propagation spans the interface is related to the particle modes of the field theory, and how the interfacial tension is expressed in terms of the particle density along the string. We then determined the order parameter and energy density profiles, and exhibited the complete agreement of these analytical results with the Monte Carlo simulations that we performed. 

The analytic derivation was performed within the field theory that describes the scaling limit of the three-dimensional Ising model in its broken phase. As usual, this limit is described by the $\phi^4$ field theory in the vicinity of its nontrivial renormalization group fixed point \cite{ID}. We exploited the particle description of this field theory, in which the particles describe the near-critical excitation modes. We showed that in the large $R$ limit that we considered the interfacial fluctuations are produced by particles that are in average largely separated, and then weakly interacting. This allowed us to obtain the exact large $R$ results (\ref{mag_profile}) and (\ref{eps_profile}), in which the information (critical exponents and amplitudes) associated to the nontrivial fixed point is contained in the magnetization $M$ and correlation length $\xi$ as specified by (\ref{xi_scaling}) and (\ref{mag_scaling}). We could then rely on the numerical values of the critical data available in the literature to perform the parameter-free comparison between analytic and Monte Carlo results shown in Figures~\ref{mag_plot} and \ref{eps_plot}.

The theoretical derivation shows that the interface exhibits Gaussian fluctuations that are not due to displacements of the interface as a whole (which would require an infinite amount of energy), but to localized excitations that, at leading order in $1/R$, involve single-particle modes\footnote{Multi-particle modes yielding subleading terms in $1/R$ can also be derived from (\ref{vPhi0}).}. These excitations propagate in the $(2+1)$-dimensional space (both momentum components $p_x$ and $p_y$ are non-zero), but the configurational average distributes them along the surface in such a way to finally yield the translational invariance of the profiles in the $y$ direction required by the boundary conditions. 

This mechanism, which involves the connectedness structure of the matrix elements of local fields on particle states, effectively implements a form of dimensional reduction in the large $R$ limit of the configurational average. This is why the magnetization profile (\ref{mag_profile}) is analogous to that in two dimensions, i.e. in absence of the $y$ axis in Figure~\ref{geometry}. The profile in two dimensions was obtained from the lattice solution of the Ising model in \cite{Abraham} (see also \cite{Abraham_review}), and more recently in field theory in \cite{DV}. The dimensional interplay holds up to an important difference: the factor $\sqrt{2}$ in (\ref{eta}) is absent in two dimensions. The origin of this difference is easy to understand in field theory. In two dimensions the particle modes of the Ising model below $T_c$ have a topological nature -- they are kinks \cite{Ryder} -- and the spin field couples only to topologically neutral states, of which the kink-antikink state is the lightest one\footnote{This corresponds to the peculiar fact that the leading singularity in momentum space of the spin-spin correlation function of the two-dimensional Ising model below the critical temperature is a branch cut rather than a pole (see \cite{McW}).} (see \cite{immf}). This is why in two dimensions the relation (\ref{xi}) is replaced by $\xi=1/2m$. It follows that in three dimensions the variance of the interfacial fluctuations expressed in terms of $\xi$ -- the measurable length scale of the statistical system -- is half of that in two dimensions.
  
The emergence of these mechanisms implies, in particular, the relevance in three dimensions of results recently obtained in two dimensions. These include those of \cite{DV} for the relation between subleading corrections in $1/R$ and the internal structure of the interface, those of \cite{bubbles,localization} for interfacial wetting \cite{Dietrich}, those of \cite{DS1,wedge,wedge2} for the effects of system geometry, and those of \cite{long_range} for the long range correlations induced by the presence of the interface. The detailed investigation of these points will provide relevant directions of further development. 

In the realm of mathematically rigorous results, the three-dimensional Ising model with the boundary conditions of Figure~\ref{geometry} has been constantly studied (see \cite{GL} and references therein) for sufficiently low temperatures (lower than the roughening temperature $T_r$) since the proof of the "rigidity" of the interface in this regime \cite{Dobrushin}. In two dimensions, several properties of Ising interfaces have been proved in recent years, for $T<T_c$ in the Ornstein-Zernike framework (see \cite{Velenik} and references therein), and for $T=T_c$ \cite{Smirnov} in the framework of Schramm-Loewner evolution (SLE) \cite{Schramm}. Our results may stimulate the mathematically rigorous investigation of the separation of phases in the three-dimensional Ising model for $T\to T_c^-$.


\vspace{1cm}
\noindent \textbf{Acknowledgments.} 
AS thanks SISSA for hospitality during the final stages of this work.

\end{document}